\documentclass[aip, amsmath, amssymb, reprint]{revtex4-2}  
\usepackage{graphicx}
\usepackage{dcolumn}
\usepackage{bm}
\usepackage[utf8]{inputenc}
\usepackage[T1]{fontenc}
\usepackage{mathptmx}
\usepackage{nccmath} 
\usepackage{hyperref} 
\hypersetup{colorlinks=true, citecolor=blue, urlcolor=blue, linkcolor=blue} 

\begin{document}

\title{Chimera states and frequency clustering in systems of \\ coupled inner-ear hair cells}

\author{Justin Faber}
\email{faber@physics.ucla.edu}
\affiliation{Department of Physics \& Astronomy, University of California, \\ Los Angeles, California 90095, USA.}
 
\author{Dolores Bozovic}
\email{bozovic@physics.ucla.edu.}
\affiliation{Department of Physics \& Astronomy, University of California, \\ Los Angeles, California 90095, USA.}
\affiliation{California NanoSystems Institute, University of California, \\ Los Angeles, California 90095, USA.}

\date{\today}

\begin{abstract}
Coupled hair cells of the auditory and vestibular systems perform the crucial task of converting the energy of sound waves and ground-borne vibrations into ionic currents. We mechanically couple groups of living, active hair cells with artificial membranes, thus mimicking \textit{in vitro} the coupled dynamical system. We identify chimera states and frequency clustering in the dynamics of these coupled nonlinear, autonomous oscillators. We find that these dynamical states can be reproduced by our numerical model with heterogeneity of the parameters. Further, we find that this model is most sensitive to external signals when poised at the onset of synchronization, where chimera and cluster states are likely to form. We therefore propose that the partial synchronization in our experimental system is a manifestation of a system poised at the verge of synchronization with optimal sensitivity.
\end{abstract}

\maketitle

\begin{quotation}
Our inner ear relies on internal nonlinearities and active, energy-consuming processes in order to detect faint sounds and comprehend speech in noisy environments. Identifying the dynamical states and mechanisms that this system utilizes to achieve such remarkable sensitivity is a long-standing open question. In this study, we identify two forms of partial synchronization in networks of coupled, living hair cells. Partial synchronization has been observed in other biological systems such as the abnormal electrical oscillations in cardiac myocytes,\citep{SATO09} and electrocorticography recordings preceding epileptic seizures.\citep{ANDRZEJAK16, LAINSCSEK19} In these examples, partial synchronization is an undesirable state. However, our experiments and simulations suggest that the inner ear may rely on partial synchronization in order to optimize its ability to detect weak signals.
\end{quotation}

\begin{figure*}[t!]
\includegraphics[width=\textwidth]{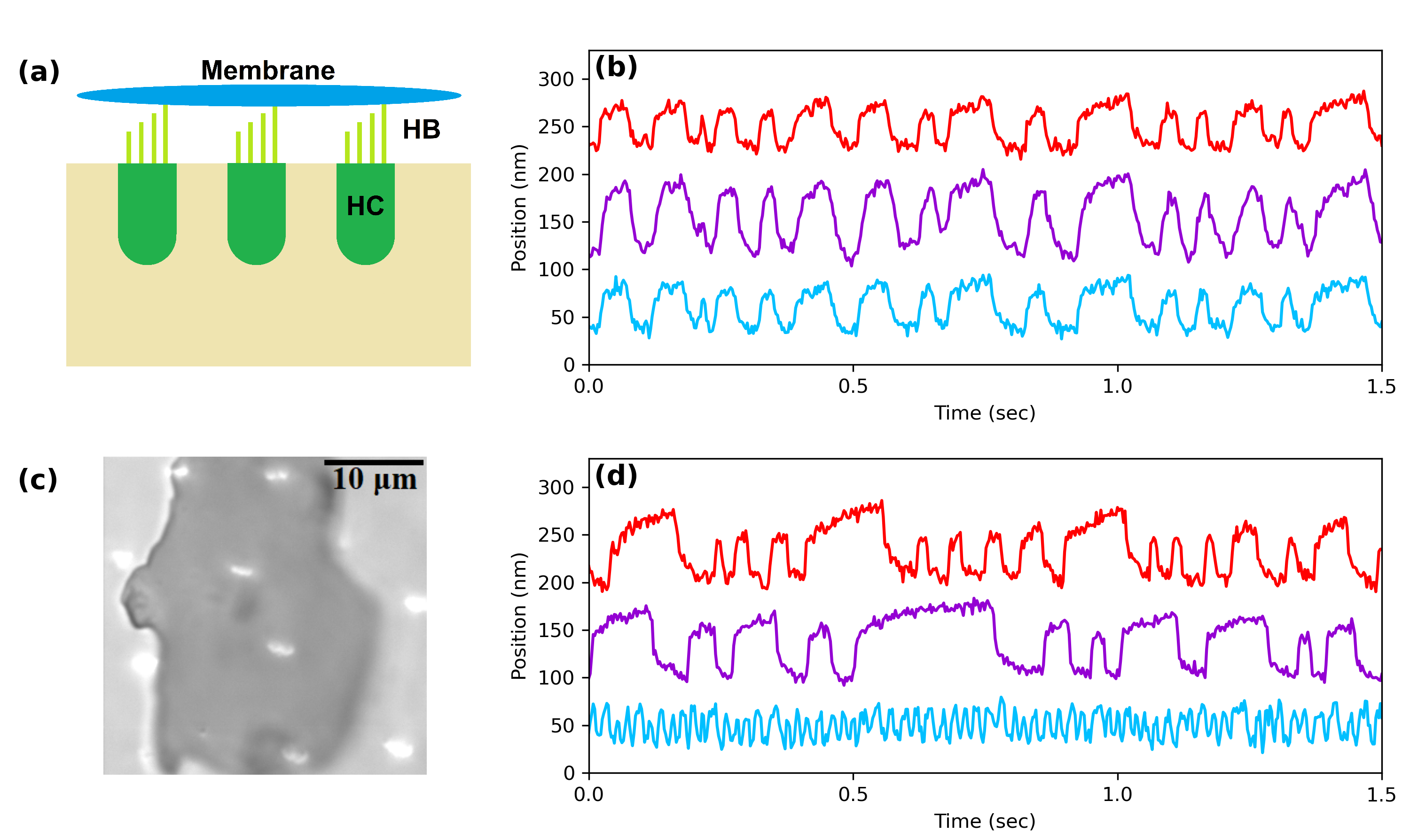}
\caption{(a) Illustration of the experimental system from a side view, displaying the hair cells (HC), hair bundles (HB), and an artificial membrane. (b) Time traces of three spontaneously oscillating hair bundles coupled by an artificial membrane. (c) Top-down images of a biological preparation. The hair bundles appear as white ovals, and the shadow cast by the transparent artificial membrane can be seen in the center of the image. (d) Time traces of the three hair bundles in (b) after removal of the artificial membrane.}
\label{Fig1}
\end{figure*}

\begin{figure*}[t!]
\includegraphics[width=\textwidth]{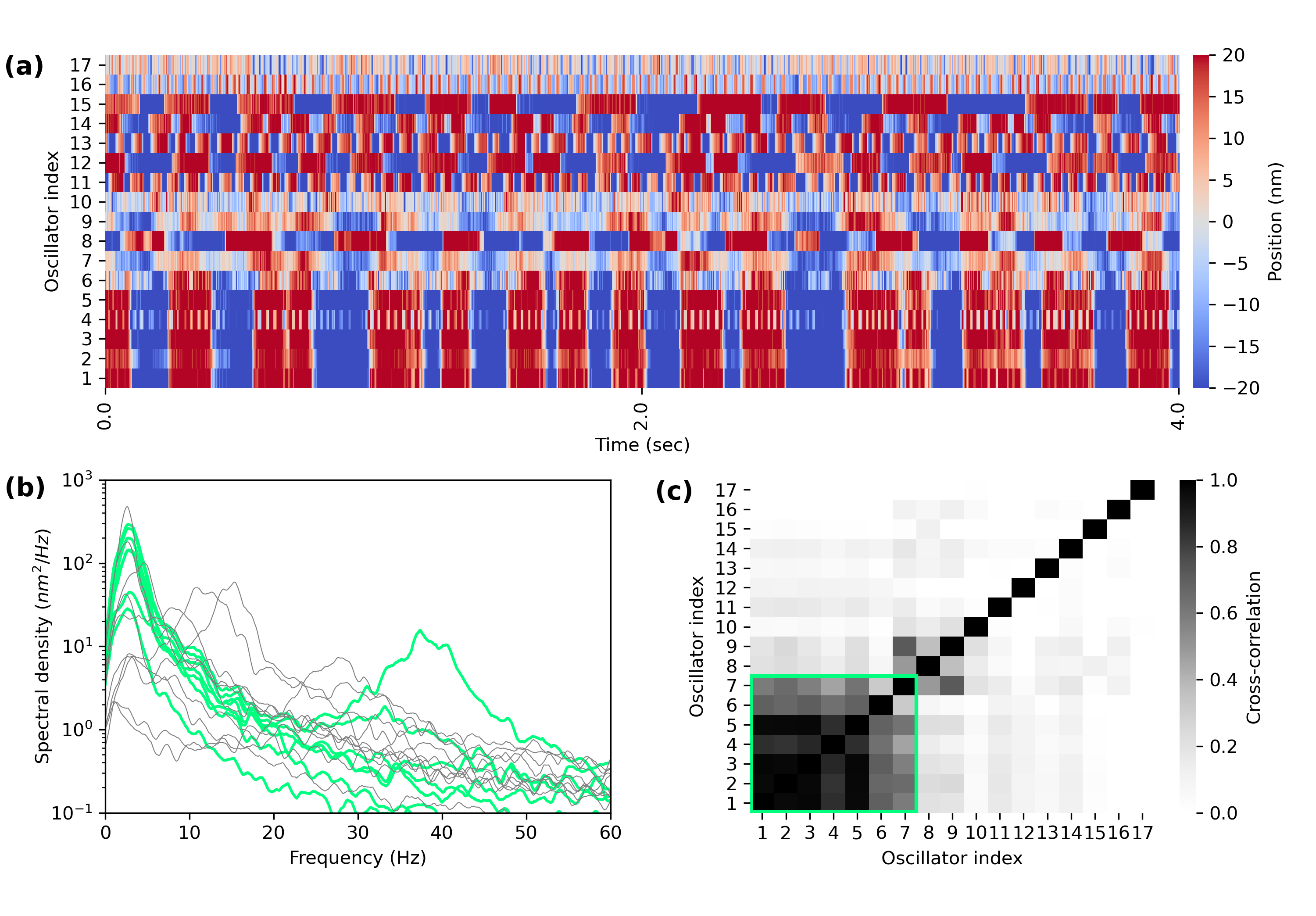}
\caption{(a) Space-time plot of 17 spontaneously oscillating hair bundles coupled with an artificial membrane, exhibiting a chimera state. Oscillators 1-7 are synchronized while the others oscillate incoherently. (b) Power spectra of the 17 coupled oscillators. The 7 synchronized hair bundles are plotting with thick, green curves. (c) Correlation matrix of the coupled hair bundles. The synchronized oscillators are highlighted in the green box.}
\label{Fig2}
\end{figure*}

\begin{figure*}[t!]
\includegraphics[width=\textwidth]{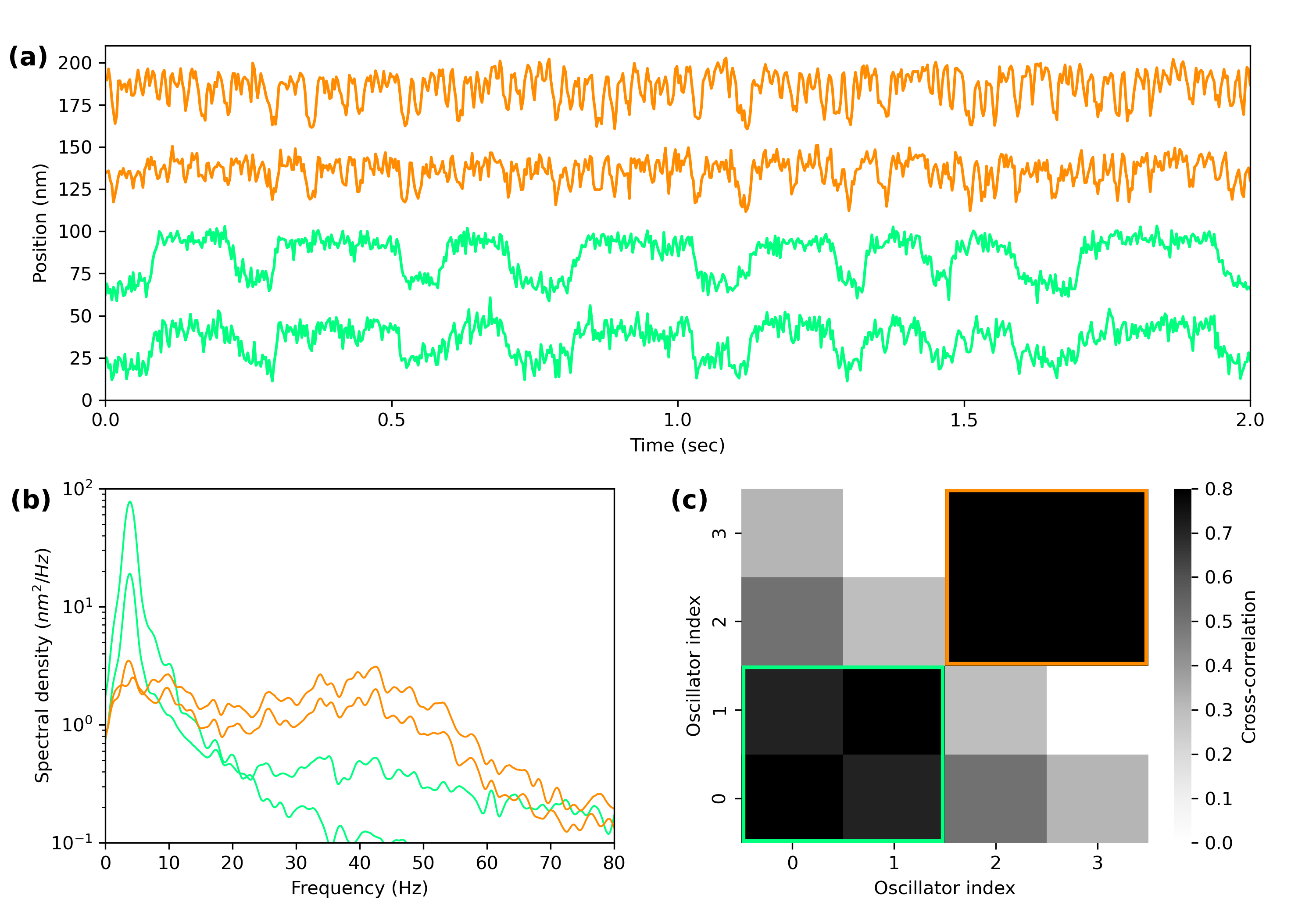}
\caption{(a) Time traces of four spontaneously oscillating hair bundles coupled with an artificial membrane, exhibiting frequency clustering. Orange (top two) and green (bottom two) traces correspond to the high and low frequency clusters, respectively. (b) Power spectra of the four coupled oscillators, plotting in colors corresponding to the traces in (a). (c) Correlation matrix of the coupled hair bundles. The synchronized oscillators are highlighted in colors corresponding to the colors of their traces.}
\label{Fig3}
\end{figure*}

\begin{figure*}[t!]
\includegraphics[width=\textwidth]{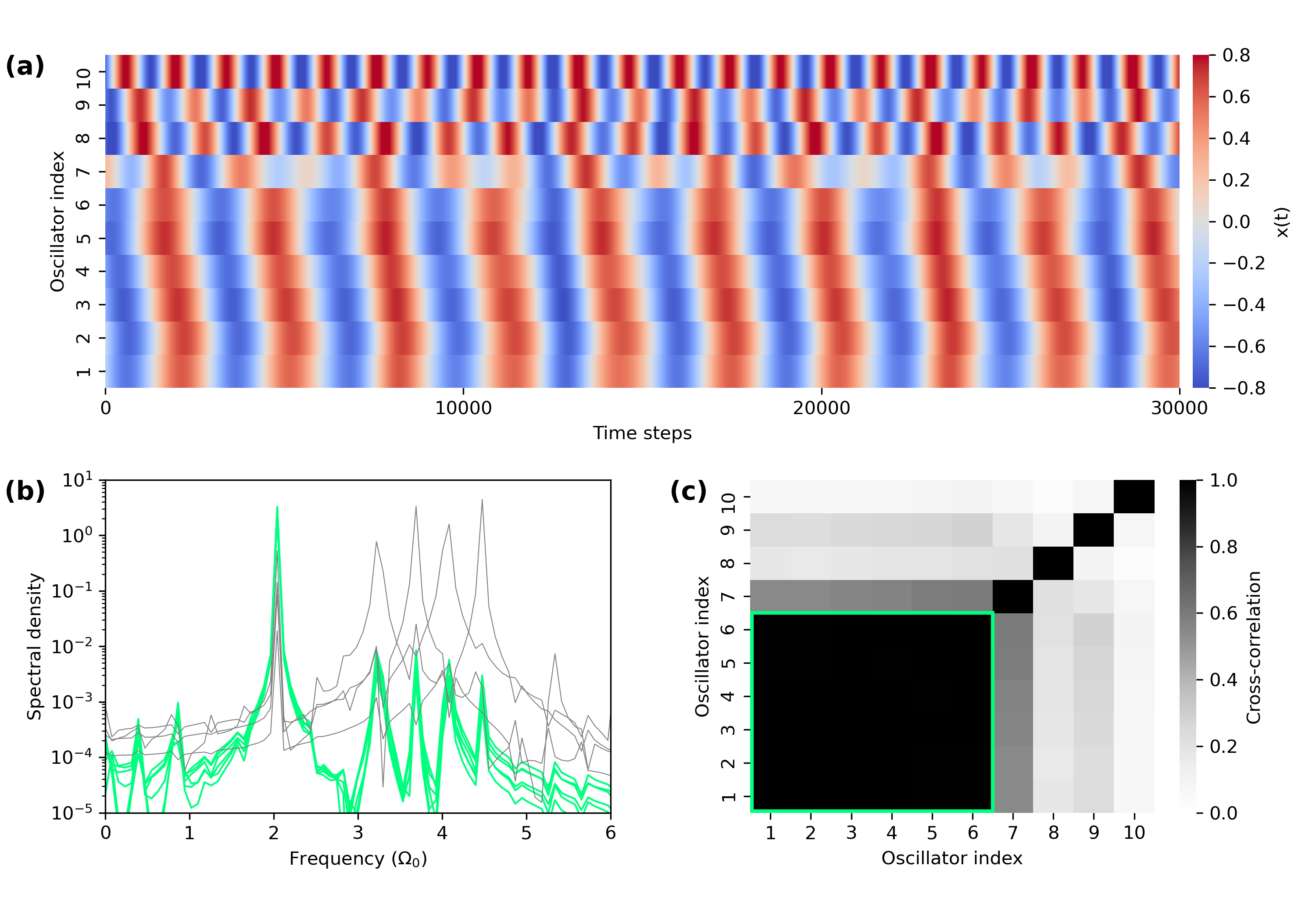}
\caption{(a) Space-time plot of the numerical model exhibiting a chimera state. Oscillators 1-6 synchronize while the others oscillate incoherently. (b) Power spectra of the 10 oscillators, with the synchronized ones plotted with thick, green curves. (c) Correlation matrix of the system. The synchronized oscillators are highlighted in the green box. For all panels, the $k_j$ values were selected from a uniform distribution spanning 0.2--3.0. The simulations were carried out with time steps of $dt=0.001$, $\beta_j = 0$, and no noise or external forcing.}
\label{Fig4}
\end{figure*}

\begin{figure*}[t!]
\includegraphics[width=\textwidth]{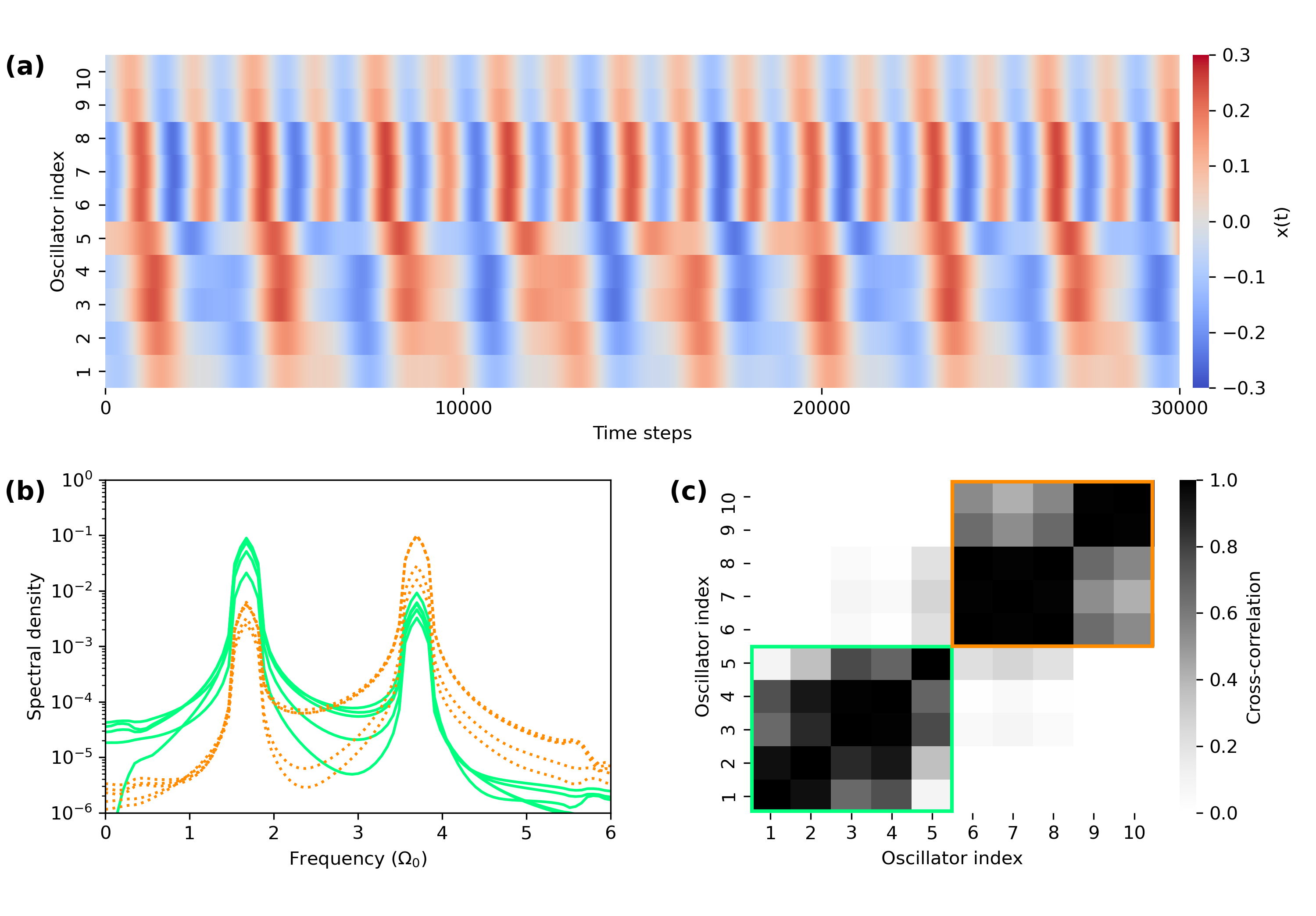}
\caption{(a) Space-time plot of the numerical model exhibiting frequency clustering. (b) Power spectra of the 10 oscillators, with five oscillators in each of the two clusters, as indicated by the solid-green (low frequency) and dotted-orange (high frequency) curves. (c) Correlation matrix of the system. The two clusters are highlighted with colors corresponding to that of their spectral density curves. Simulations were performed with no noise or external forcing, and with time steps of $dt=0.001$. The coupling strengths were all equal, $k_j = k = 1.5$, and the $\beta_j$ values were selected from uniform distributions spanning -1--0 for the first five oscillators and 0--1 for the last five oscillators.}
\label{Fig5}
\end{figure*}

\begin{figure*}[t!]
\includegraphics[width=\textwidth]{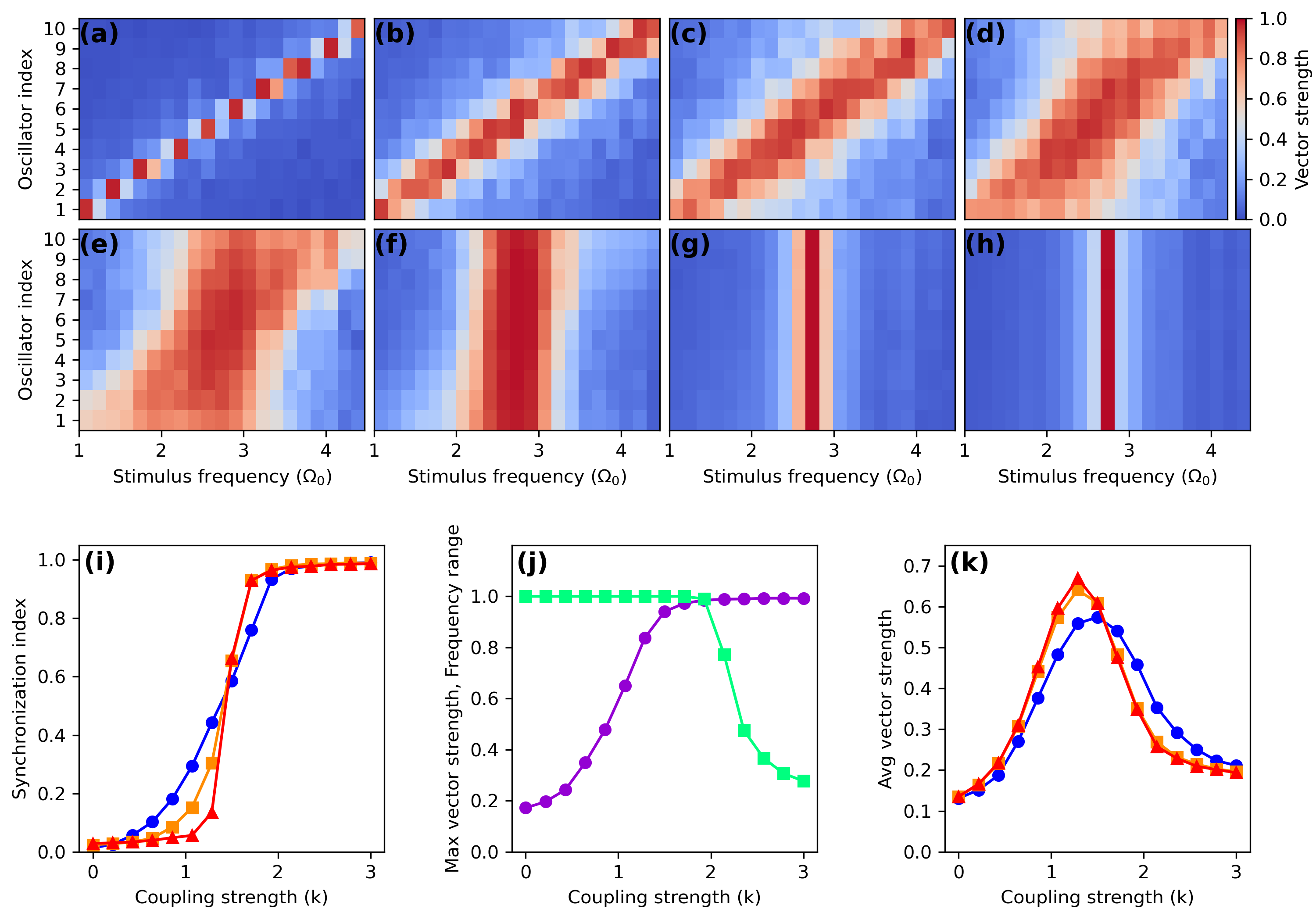}
\caption{(a-h) Sensitivity maps of the numerical model in response to a sinusoidal stimulus with amplitude $f_0 = 0.2$ and noise strength $D=0.01$ for coupling strengths $k_j = k = 0$, $0.75$, $1$, $1.25$, $1.5$, $2$, $3$, and $5$, respectively. (i) Synchronization index as a function of the coupling strength, showing the transition to coherence. (j) Trade-off between the oscillator-average vector strength at the most sensitive stimulus frequency (purple circles) and the fraction of stimulus frequencies for which an oscillator responds with a vector strength of at least 0.2 (green squares). (k) Vector strength averaged across all oscillators and all stimulus frequencies. For (i) and (k), blue circles, orange squares, and red triangles represent system sizes of 10, 100, and 1000 coupled oscillators, respectively.}
\label{Fig6}
\end{figure*}

\section{Introduction}

The auditory and vestibular systems are extraordinary signal detectors. These end organs can reliably detect sound and mechanical vibrations that induce displacements as small as a few angstroms, comparable to or below the amplitude of motion caused by thermal fluctuations in the surrounding fluid.\citep{Hudspeth14} These sensory systems also exhibit remarkable temporal resolution, frequency selectivity, and dynamic range of detection. How these biological sensors achieve their signal detection properties is a long-standing open question, and the physics of hearing remains an active area of research.\citep{REICHENBACH14}

Mechanical detection of sound waves, vibrations, and accelerations is performed by hair cells. These specialized cells are named after the rod-like stereovilli that protrude from their apical surfaces. The cluster of inter-connected stereovilli is named the hair bundle and performs the essential task of transducing the mechanical energy of sound into electrical signals that take the form of ionic currents into the cell.\citep{LEMASURIER05, VOLLRATH07, OMAOILEIDIGH19} A perturbation caused by sound or acceleration results in a deflection of the hair bundles and an increase in the tension of the tip links that connect adjacent rows of stereovilli. A change in tension of the tip links modulates the open probability of the transduction channels that are embedded at the tops of the stereovilli and connected to the tip links. 

Auditory detection has been shown to require an active, energy-consuming process in order to achieve such remarkable signal detection.\citep{HUDSPETH08} This active process manifests itself in a number of phenomena, including the appearance of autonomous motion of the hair bundles, observed \textit{in vitro} in several species.\citep{BENSER96, MARTIN03, CRAWFORD85} These spontaneous oscillation have amplitudes well above the noise induced by thermal fluctuations, and they have been shown to be active, as they violate the fluctuation dissipation theorem.\citep{MARTIN01} The role of these spontaneous oscillations is not yet fully understood, but prior studies have suggested that they could be utilized as an amplification mechanism for weak signals.\citep{MARTIN99} These spontaneous oscillations also serve as a probe for studying the active cellular mechanics underlying auditory detection.

Another manifestation of this active process is the spontaneous emission of sound, observed \textit{in vivo} in many species.\citep{KEMP79} These spontaneous otoacoustic emissions (SOAEs) exhibit several sharp peaks in their power spectra and are metabolically sensitive, indicating an underlying energy-consuming process. Although SOAEs serve as a diagnostic for hearing-related disorders in humans, there is currently no consensus on the mechanism responsible for generating them.\citep{SHERA04, BERGEVIN15, ROONGTHUMSKUL19} One theory suggests that they arise from frequency clustering of actively oscillating coupled hair cells.\citep{VILFAN08, FRUTH14}

\textit{In vivo}, hair bundles are attached to overlying structures, which provide coupling between the individual active oscillators. The strength and extent of the coupling varies across species and organs.\citep{OMAOILEIDIGH19} In the bullfrog sacculus, several thousand hair cells are coupled together by the otolithic membrane. The sacculus is responsible for detecting low-frequency ground-borne and airborne vibrations. In contrast to auditory organs, the sacculus does not display a high degree of frequency selectivity, nor any tonotopic organization of the hair cells: there is no correlation between the characteristic frequencies of the hair cells and their location in the sensory epithelium.\citep{SMOTHERMAN00} It does, however, demonstrate extreme sensitivity of detection.\citep{NARINS84} We previously demonstrated that, despite frequency dispersion as large as five-fold, groups of coupled hair bundles can fully synchronize.\citep{FABER21} Our experimental and theoretical studies indicated that the presence of chaotic dynamics in individual oscillators enhances synchronization in the coupled system and allows for highly sensitive detection, even in the presence of biological levels of noise. Other work has proposed that systems of coupled bundles can also exhibit an amplitude death regime, with quenching caused by strong coupling and significant frequency dispersion.\citep{KIM14}

In this study, we identify two additional dynamical states that can occur in networks of coupled hair bundles and explore their potential role in the detection capabilities of the auditory system. The first is the chimera state, defined as a system in which a subset of the coupled oscillators shows mutual synchronization, while the rest oscillate incoherently.\citep{PANAGGIO15} Previously it had been believed that identical oscillators, coupled through a mean field, could occupy only two dynamical states: full synchronization or incoherence. This assumption was disproven by the observation of chimeras, first seen in numerical simulations of identical oscillators.\citep{KURAMOTO02, ABRAMS04} As the presence of chimera states depends strongly on the initial conditions of the dynamical system, it was believed that they were too unstable to be observed in an experimental system. However, a decade after their discovery in numerical simulations, chimera states were observed in coupled chemical oscillators\citep{TINSLEY12} and in coupled-map lattices.\citep{HAGERSTROM12} Chimera states were also shown to arise from heterogeneity in the parameters of the coupled oscillators.\citep{PANAGGIO15} As hair cells inherently possess heterogeneity in their size, structure, and time scales of ion-channel dynamics, systems of coupled hair bundles can support chimera states. We here demonstrate signatures of chimeras in experimental recordings obtained from hybrid preparations, in which artificial coupling structures are interfaced with live hair cells. We further explore their potential role in signal detection, with theoretical models that simulate hair bundle dynamics. 

The second dynamical phenomenon explored in this study is the occurrence of cluster states, another form of partial synchronization in a coupled system, in which each oscillator synchronizes with one of several clusters. We identify states of frequency clustering in our \textit{in vitro} experiments, lending support to the theory that SOAEs may be generated by frequency clustering of actively oscillating hair bundles. Both types of partial synchronization, chimeras and cluster states, can be reproduced by our simple numerical model of hair cell dynamics, with the introduction of heterogeneity in the set of parameters.

Lastly, we use our numerical model to test its sensitivity to external signals, when the system resides in different dynamical regimes. The system is most frequency selective when poised in the regime of strong coupling, where all oscillators synchronize. However, consistent with a previous numerical study,\citep{WANG17} we find that the sensitivity of the system is maximized in the regime of intermediate coupling strength, near the onset of synchronization, in which chimera and cluster states are likely to arise. As hair cells have been shown to utilize a number of adaptive mechanisms, we therefore speculate that the coupled systems within auditory and vestibular end organs may poise themselves at the onset of synchronization in order to optimize their sensitivity to weak, external signals.

\section{Chimera and Cluster states \textit{in vitro} }

We explore the occurrence of partial synchronization in experimental systems of coupled hair cells. To introduce artificial coupling, we use mica flakes and attach them to the tops of groups of hair-cell bundles, following our previously developed methods.\citep{FABER21} The artificial membranes are dispersed atop the bundles by introducing them into the endolymph solution, which bathes the apical surface of the sensory epithelium. The mica membranes are thin and transparent, allowing for precise imaging and tracking of the motion of the underlying hair bundles. Further, the mica flakes only minimally modify the mass and drag of the system, allowing us to explore the dynamics of a wide range of system sizes. 

As mentioned previously, the bullfrog sacculus is not tonotopically organized, and adjacent hair bundles exhibit up to tenfold differences in their frequencies of spontaneous oscillation. Despite this large frequency dispersion, we find that groups of neighboring hair bundles routinely synchronize upon coupling by the artificial membranes (Fig. \ref{Fig1}). We characterize the degree of synchronization by calculating the cross-correlation coefficient (Eq. \ref{cross_cor}) between each pair of hair bundles in the system.

Due to the differences in heights of the stereovilli, not every hair bundle within a group makes contact with the artificial membrane above it. Therefore, we define a threshold to determine which hair bundles are coupled to a network of others. To find this threshold, we first calculate the cross-correlation coefficient between many unique pairs of uncoupled hair bundles, with no artificial membranes in the vicinity. The distribution of cross-correlation coefficients is centered around $0$ and has a standard deviation $\approx 0.02$ (Fig. \ref{Fig7}). We then set the threshold to be $0.1$, more than five standard deviations above the mean and hence unlikely to occur by chance without coupling.

In addition to fully synchronized states, we also observe cases of partial synchronization. We use several techniques to characterize these states. First, we generate space-time plots, where the traces of all the oscillators in the coupled system are plotted as a function of time and the amplitude is represented by color, providing a visual observation of synchronization between the oscillators (Figs. \ref{Fig2}a, \ref{Fig3}a). Next, we view the power spectra of the oscillators to confirm that the dominant peaks align at a common frequency for the synchronized portion of the chimera states (Fig. \ref{Fig2}b) and that multiple common peaks are present for the oscillators of the cluster states (Fig. \ref{Fig3}b). Finally, we plot the correlation matrices of all of the traces within the coupled systems, in order to give another visual representation of the partial synchronization states. Chimera states contain one group of oscillators with large cross-correlation coefficients between each pair, while all other cross-correlation coefficients are low (Fig. \ref{Fig2}c). In contrast, cluster states contain multiple groups, where oscillators within a group are strongly correlated with each other but not with those outside of the group (Fig. \ref{Fig3}c).

\section{Numerical Model of coupled hair cells}

The dynamics of the $j^{th}$ oscillator in the coupled system are described using the complex variable, $z_j(t) = x_j(t) + iy_j(t)$, and are assumed to be governed by the normal form equation for the supercritical Hopf bifurcation,

\begin{eqnarray}
\frac{dz_j}{dt} = (\mu + i\omega_j)z_j - (1 + i\beta_j)|z_j|^2 z_j + k_j(\bar{z} - z_j) \nonumber \\ + \eta_j(t) + F(t).
\label{eq:Hopf}
\end{eqnarray}

\noindent This simple model reproduces many of the experimentally observed phenomena of hair-cell dynamics, such as the autonomous oscillations and the compressive nonlinear response to external signals.\citep{EGUILUZ00, KERN03} The real part of $z_j(t)$ represents the hair bundle position, while the imaginary part reflects internal parameters of the cell and is not assigned a specific, measurable quantity. $\mu$ controls the proximity to the Hopf bifurcation, and $\omega_j$ represents the natural frequency at this bifurcation ($\mu=0$) in the absence of coupling. $\beta_j$ characterizes the degree of nonlinearity and controls the level of nonisochronicity of the oscillator.\citep{FABER19a, ROONGTHUMSKUL21} In the absence of coupling, and for $\mu > 0$, the system exhibits limit cycle oscillations at radius $\sqrt{\mu}$ and frequency $\Omega_j = \omega_j - \beta_j\mu$. We set the frequency dispersion in our model to approximate that of our experimental data. The limit cycle frequencies are hence uniformly spaced from $\Omega_1 = 1$ to $\Omega_N = 2\sqrt{5}\approx4.47$. We select an irrational number to avoid spurious mode-locking between oscillators. We set the control parameter to be $\mu=1$ throughout the study, poising the system deep into the oscillatory regime.

The system is subject to external real-valued forcing, $F(t)$, representing acoustic stimulus or linear acceleration, both of which elicit deflection of the hair bundles in the sacculus. Each oscillator is subject to independent, additive white Gaussian noise, $\eta_j(t)$, with independent real and imaginary parts: $\langle Re(\eta_j(t))Re(\eta_j(t')) \rangle  = \langle Im(\eta_j(t))Im(\eta_j(t')) \rangle = 2D \delta (t-t')$, where $D$ is the noise strength of the system.

The dynamics of this system occur at low Reynolds number,\citep{CIGANOVIC19} and we have previously shown that the drag of the artificial membranes is small in comparison to that of the entire coupled system.\citep{FABER21} Further, the mica flakes exhibit little compliance, as we observe coupling and synchronization between pairs of hair bundles with large spacial separation. For these reasons, we have chosen to model the system with mean-field coupling, where each oscillator is weighted by its degree of attachment to the artificial membrane, $k_j$. The weighted mean field then takes the form,

\begin{eqnarray}
\bar{z} = \frac{1}{N} \sum_{j=1}^N k_j z_j,
\label{eq:Hopf}
\end{eqnarray}

\noindent where $N$ is the number of oscillators in the system.

\section{Chimera and Cluster States in the numerical simulations of hair cell dynamics}

To explore the dynamic states that can occur in the system of coupled hair cells, we perform numerical simulations based on the theoretical model described above. We vary parameters over a physiologically plausible range, to reproduce the dynamical states observed in the experimental system and determine their potential mechanisms. As mentioned earlier, chimera states can arise from heterogeneity of the model parameters. In the biological system, it is unlikely that the level of attachment to the membrane is identical for all oscillators; hence, we randomly select each $k_j$ value from a uniform distribution spanning 0.2--3.0. This heterogeneity tends to produce the chimera state in our simulations, as only some of the oscillators synchronize, while others oscillate incoherently (Fig. \ref{Fig4}). 

Next, we explore the effects of introducing dispersion into the selection of the $\beta_j$ parameters, which control the degree of nonisochronicity in individual oscillators. We have previously shown that this parameter, which renders the oscillation frequency of an oscillator dependent on its oscillation amplitude,  can lead to chaotic dynamics in the presence of noise.\citep{FABER19b} Further, we demonstrated that it can enhance synchronization in a system of coupled nonlinear oscillators, as it allows for a greater shifts in the innate frequencies of oscillation.\citep{FABER21} Here we show that random dispersion in this parameter can also result in multiple frequency clusters (Fig. \ref{Fig5}). We find that the system forms a 2-cluster state, where oscillators with positive and negative $\beta_j$ values form separate clusters. The clustering results from coupling, which tends to restrict the dynamics to a smaller region of phase space, thus reducing the amplitude. Since the sign of $\beta_j$ determines whether an oscillator's frequency increases or decreases with amplitude reduction, two distinct frequencies emerge, forming stable clusters.

\section{Optimization of Signal Detection}

To achieve reliable signal detection, a group of coupled detectors may utilize synchronization. The inherent noise of each component is thus averaged out, and the signal-to-noise ratio (SNR) increases with increasing number of detectors.\citep{DIERKES12} The drawback to complete synchronization is that the system is then sensitive to only a small range of frequencies surrounding the characteristic frequency. Therefore, total synchronization would be an unfavorable state for the groups of coupled hair cells in the auditory systems, as they are responsible for detecting frequencies that span several octaves. We therefore propose that these systems may utilize a low degree of synchronization in order to improve the SNR without compromising the frequency range of detection.

To visualize this inherent trade-off as a function of the coupling strength, we construct maps that display the sensitivity of every oscillator to a wide range of stimulus frequencies (Fig. \ref{Fig6}a-h). We characterize the degree of phase locking, by calculating the vector strength,

\begin{eqnarray}
v_{ij} = \sqrt{ \langle \sin(\phi_i - \phi_j) \rangle^2 + \langle \cos(\phi_i - \phi_j) \rangle^2},
\label{eq:Hopf}
\end{eqnarray}

\noindent where $\phi_i$ and $\phi_j$ are the phases of two time series and the angle brackets denote the time average. To quantify the sensitivity, we calculate the vector strength between the stimulus waveform and the response of an oscillator. 

To characterize the degree of synchronization within the coupled system we calculate the average vector strength between all pairs of oscillators in the absence of stimulus. This synchronization index is 1 for perfectly synchronized oscillators and approximately 0 for incoherent motion. We see that the transition to synchronization occurs at a coupling strength around $k_j = k = 1.5$ and becomes more abrupt for larger system sizes (Fig. \ref{Fig6}i).

The sensitivity maps show the strongest response at intermediate levels of coupling strength, near the onset of synchronization. If the coupling is too weak, the dynamics are incoherent, and the oscillators are more susceptible to noise. However, if the coupling is too strong, the system is limited to detecting only a small range of frequencies. To see this trade-off more explicitly, we average the vector strengths over all oscillators and take the maximum across all stimulus frequencies (maximum oscillator-average vector strength). We plot this measure as a function of coupling strength, along with the fraction of stimulus frequencies in which at least one oscillator has a vector strength above 0.2 (Fig. \ref{Fig6}j). This trade-off between maximum vector strength and frequency range of detection produces a peak in the average vector strength across all stimulus frequencies and detectors (Fig. \ref{Fig6}k). These results suggest that a coupled system responsible for detecting a wide range of frequencies will achieve optimal performance when poised at the onset of synchronization. 

Vestibular end organs display varying degrees of coupling between hair cells, likely involving the response of multiple oscillators to achieve reliable signal detection. The bullfrog sacculus is innervated in a way that supports this assumption, with afferent fibers synapsing onto multiple hair cells.\citep{SMOTHERMAN00} Secondly, to achieve reliable detection, the system should be sensitive to different frequencies, as the airborne and ground-borne vibrations of interest contain energy distributed across a range of low frequencies. We therefore propose that the sacculus is poised at the onset of synchronization in order to optimize signal detection.

\section{Discussion}

The auditory and vestibular systems have provided a testing ground for concepts from bifurcation theory and nonlinear dynamics.\citep{EGUILUZ00, CAMALET00, NEIMAN11, JI18} These sensory systems serve different purposes but all rely on hair cells to perform detection of sound, vibration, or acceleration. Active hair cells of these sensory systems have displayed Hopf bifurcations, saddle-node on an invariant circle (SNIC) bifurcations,\citep{FREDRICKSON12} and the quasiperiodic transition to chaos.\citep{FABER18} The dynamics of hair cells have been described by limit cycles, stable fixed points, chaotic attractors, and amplitude-death states. How these dynamical states shape the response of the full system to external signals remains an open question.

As these sensory systems impose different requirements on the sensitivity, frequency selectivity, temporal resolution, and dynamic range of detection, the various organs may have developed different dynamical regimes in which to reside, in order to achieve the signal detection properties of interest. Individual hair cells that comprise these systems have been shown to be versatile, displaying different response characteristics under different mechanical loads or other perturbations.\citep{SALVI15} For example, while the bullfrog sacculus is not a frequency-selective organ, when hair cells within it are subject to appropriate experimental manipulation, they were shown to be capable of frequency-selective detection expected for auditory organs. Hence, we expect that differences in the detection properties of the sensory organs lie not only in different properties of individual cells, but also in the coupling conditions and emergent dynamical states of the full system. It is therefore important to understand the different dynamical states that these coupled oscillators can exhibit, in order to understand the full range of signal detection properties that the sensory systems can display.

In the present work, we observe two dynamical states that, to the best of our knowledge, have not been previously observed in auditory or vestibular systems. We measure the response of active hair bundles, coupled together with artificial membranes of different sizes, and obtain experimental observations of chimera states and cluster states. Both of these dynamical states can be reproduced with a simple numerical model with the inclusion of heterogeneity of the parameters. 

One of the signatures of the active process in the inner ear takes the form of otoacoustic emissions, which have been used as a probe of the auditory nonlinearities and internal dynamics \textit{in vivo}.  The mechanism of their generation by the auditory system is, however, not yet fully established. Our experimental data supports the theory that they arise from frequencies clustering of coupled active oscillators within the inner ear, as we observe frequency clustering \textit{in vitro} in small groups of mechanically coupled hair bundles. 

Both the cluster states and chimera states are forms of partial synchronization and arise at intermediate levels of coupling strength, near the onset of total synchronization. We find that the numerical model achieves the greatest sensitivity to external stimulus when poised in this regime. It exhibits a balance in the inherent trade-off between the frequency range of signal detection and the number of oscillators that phase lock to the external signals. Therefore, we propose that these partial synchronization states may occur \textit{in vivo} in systems of coupled hair bundles, if these systems are poised in the optimal regime for signal detection.

\section*{acknowledgments}
The authors gratefully acknowledge the support of NSF Biomechanics and Mechanobiology, under Grant No. 1916136. The authors thank Dr. Sebastiaan Meenderink for developing the software used for tracking hair bundle movement.

\section*{data availability}
The data that support the findings of this study are available
from the corresponding author upon reasonable request.

\appendix* 
\section{Experimental Methods}

\subsection{Biological Preparation}
Experiments were performed \textit{in vitro} on hair cells of the American bullfrog (\textit{Rana catesbeiana}). Sacculi were excised from the inner ear of the animal and mounted in a two-compartment chamber with artificial perilymph and endolymph solutions \citep{BENSER96}, mimicking the natural conditions for this tissue. Hair bundles were accessed after digestion and removal of the overlying otolithic membrane \citep{MARTIN01}. All protocols for animal care and euthanasia were approved by the UCLA Chancellor's Animal Research Committee in accordance with federal and state regulations.

\subsection{Artificial Membranes}
Mica powder was added to a vial of artificial endolymph solution, thoroughly mixed, and then filtered through several steel mesh gratings. These gratings served as filters to extract only the desired size of mica flakes, typically 20--50 $\mu$m. The solution was then pipetted into the artificial endolymph solution, above the biological preparation, causing mica flakes to settle on top of the hair bundles. The hair bundles adhered to the surface of the mica, resulting in coupling.

\subsection{Data Collection}
Hair bundle motion was recorded with a high-speed camera at frame rates between 250 and 1000 frames per second. The raw images were analyzed in MATLAB to determine the position of the center of the hair bundle in each frame. The motion was tracked along the direction of increasing stereovilli height. Typical noise floors of this technique, combined with stochastic fluctuations of the bundle position in the fluid, were 3--5 nm.

\subsection{Cross-Correlation Coefficient}

\begin{figure}[h!]
\includegraphics[width=\columnwidth]{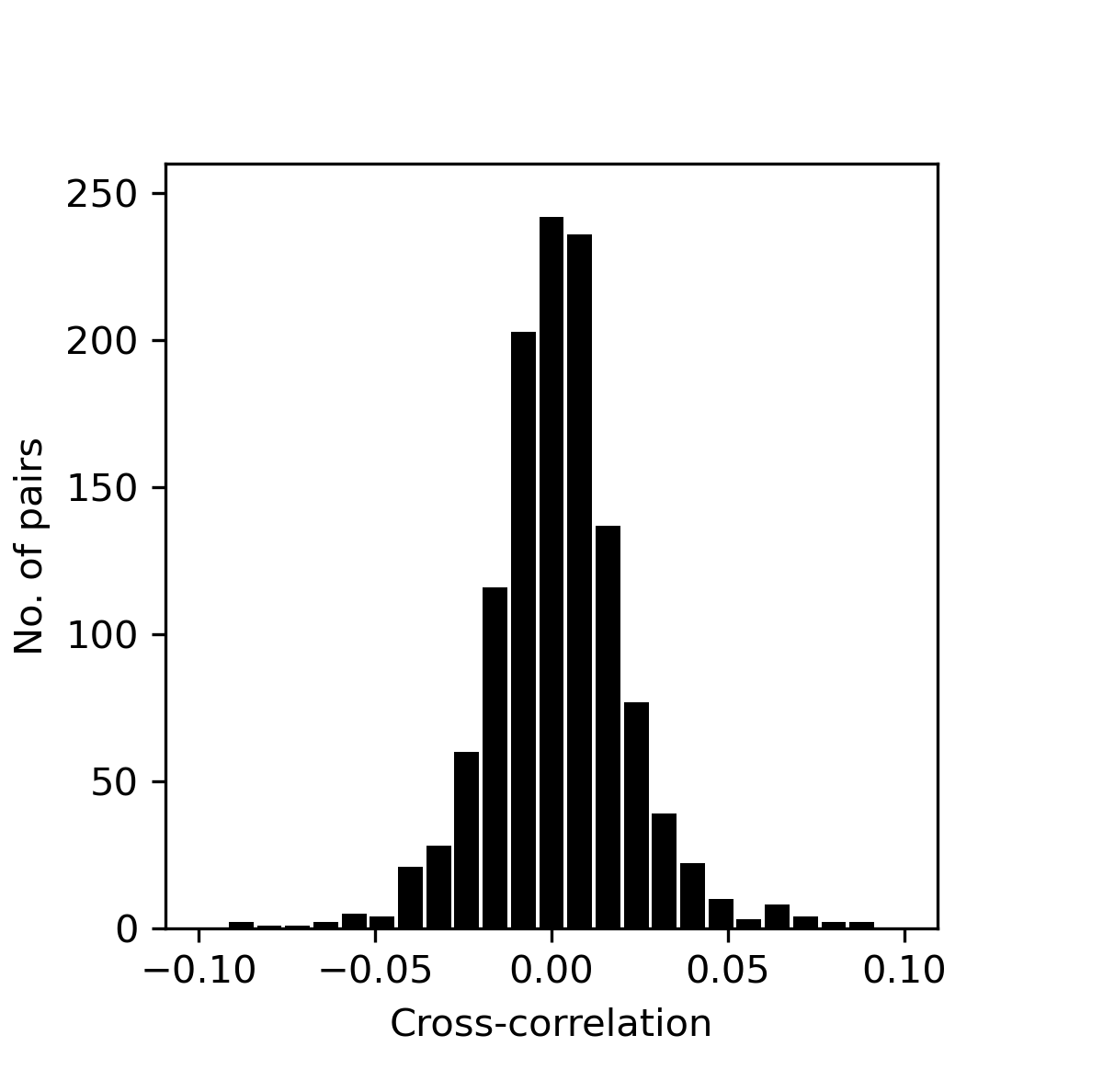}
\caption{Histogram of cross-correlation coefficients between pairs of uncoupled, spontaneously oscillating hair bundles (1225 unique pairs). The standard deviation of this distribution is $<0.02$ and no points exceed 0.1.}
\label{Fig7}
\end{figure}

We characterize synchronization between spontaneously oscillating hair bundles using the cross-correlation coefficient

\begin{eqnarray} \label{cross_cor}
C\big(x_1(t), x_2(t)\big) = \frac{\langle \tilde{x}_1(t) \tilde{x}_2(t) \rangle}{\sigma_1 \sigma_2},
\end{eqnarray}

\noindent where $\tilde{x}_1(t) = x_1(t) - \langle x_1(t) \rangle$ and $\tilde{x}_2(t) = x_2(t) - \langle x_2(t) \rangle$ represent the time traces of the motion with zero mean, $\sigma_1$ and $\sigma_2$ represent their respective standard deviations, and the angle brackets denote the time average. $C=1$ indicates perfectly correlated motion, while $C \approx 0$ indicates uncorrelated motion. We do not consider hair bundles to be coupled unless they their cross-correlation coefficient exceeds our threshold of 0.1 with other hair bundles in a network. This threshold is based on the values of the measure in the absence of coupling (Fig. \ref{Fig7}).

\newpage

\bibliography{Bibliography}
\end{document}